\documentclass[twocolumn]{webofc}
\usepackage[varg]{txfonts}   % Web of Conferences font

\def\Sw{{\em Swift}}

\newcommand\beq{\begin{equation}}
\newcommand\eeq{\end{equation}}

\woctitle{Physics at the Magnetospheric Boundary}
\newcommand{\apj}{ApJ}
\newcommand{\mnras}{MNRAS}

\newcommand{\aap}{A\&A}                   % "Astron. Astrophys."
\begin{document}
\title{X-ray and UV correlation in the quiescent emission of Cen X-4, evidence of accretion and reprocessing}

\author{F.~Bernardini\inst{1,2}\fnsep\thanks{\email{bernardini@wayne.edu}} \and
        E.~M.~Cackett\inst{1} \and
        E.~F.~Brown\inst{3}   \and
        C.~D'Angelo\inst{4}   \and
        N.~Degenaar\inst{5,6} \and
        J.~M.~Miller\inst{5}  \and
        M.~Reynolds\inst{5}  \and
        R.~Wijnands\inst{4}
}

\institute{Department of Physics \& Astronomy, Wayne State University, 666 W. Hancock St., Detroit, MI 48201, USA\and
INAF, Osservatorio Astronomico di Capodimonte, Salita Moiariello 16, 80131 Napoli, Italia\and
Department of Physics \&  Department of Physics, Michigan State University, East Lansing, MI 48824, USA\,\and
Instituut Anton Pannekoek, University of Amsterdam, Amsterdam 1098 XH, The Netherlands \and
Department of Astronomy, University of Michigan, 500 Church St, Ann Arbor, MI 48109-1042, USA\and
Hubble fellow}

\abstract{We conducted the first long-term (60 days), multiwavelength (optical, ultraviolet, and X-ray) simultaneous monitoring of Cen X-4 with daily \Sw\ observations, with the goal of understanding variability in the low mass X-ray binary Cen X-4 during quiescence. We found Cen X-4 to be highly variable in all energy bands on timescales from days to months, with the strongest quiescent variability a factor of 22 drop in the X-ray count rate in only 4 days. The X-ray, UV and optical (V band) emission are correlated on timescales down to less than 110 s. The shape of the correlation is a power law with index $\gamma$ about 0.2--0.6. The X-ray spectrum is well fitted by a hydrogen NS atmosphere ($kT=59-80$ eV) and a power law (with spectral index $\Gamma=1.4-2.0$), with the spectral shape remaining constant as the flux varies. Both components vary in tandem, with each responsible for about 50\% of the total X-ray flux, implying that they are physically linked. We conclude that the X-rays are likely generated by matter accreting down to the NS surface. Moreover, based on the short timescale of the correlation, we also unambiguously demonstrate that the UV emission can not be due to either thermal emission from the stream impact point, or a standard optically thick, geometrically thin disc. The spectral energy distribution shows a small UV emitting region, too hot to arise from the accretion disk, that we identified as a hot spot on the companion star. Therefore, the UV emission is most likely produced by reprocessing from the companion star, indeed the vertical size of the disc is small and can only reprocess a marginal fraction of the X-ray emission. We also found the accretion disc in quiescence to likely be UV faint, with a minimal contribution to the whole UV flux.
}
\maketitle
\section{Introduction}
\label{intro}

In low mass X-ray binaries (LMXBs), a compact object (a neutron star, NS, or a black hole, BH), is accreting matter from a companion star with mass smaller than that of the Sun. The physics of LMXBs during their outburst state is thought to be well established. The general properties of the quiescence to outburst cycle is explained by the disc instability model, DIM \citep[][]{cannizzo93,laso}.  
On the contrary the mechanism powering the optical, UV and X-ray emission in quiescence is still debated. The emission could be powered by residual accretion, however, the physics of accretion at low Eddington luminosity rates is far from being understood. Several models have been proposed to explain the difference in the emission from an LMXB containing a NS and a BH. However, so far, a unifying scenario which can make a clear prediction, systematically matching the spectral energy distribution (SED), from optical up to the X-ray emission, of both NS and BH quiescent LMXB, is still missing.

Despite much early observational and theoretical work on quiescent LMXBs, there are several crucial questions still remaining which only simultaneous multiwavelength campaigns can address.  For instance, is it accretion at low rates really happening onto NSs, and how exactly does it work? Where precisely does the UV emission arise from? Is the X-ray emission irradiating perhaps the inner edge of the accretion disc, leading to reprocessed UV emission, which would imply that X-ray variability is triggering UV variability?  Otherwise, are mass accretion rate fluctuations propagating inwards from the outer disc (UV), up to the NS surface (X-ray), implying that, on the contrary, UV variability is leading the X-ray variability?
In order to answer these questions, we planned a new and unique study of the NS LMXB Cen X-4. For the first time ever, we intensively monitored the source on a daily basis, for more than two months, with simultaneous optical, UV, and X-ray observation performed by \Sw. Here we present a brief summary of the main results 
of our fully comprehensive paper \cite{bernardini13} which has been presented at the conference: \textit{Physics at the Magnetospheric Boundary} (Geneve, Switzerland, 25-28 June 2013).

\section{Results}
\label{ana&res}
\subsection{Multiwavelength light curve}
\label{lightcurve}

Cen X-4 is highly variable in all energy bands. Particularly interesting is the comparison between the X-ray and the UVW1 and V bands.
The X-ray and UVW1 light curve seem to follow roughly the same pattern, showing both long term (months) and short term (days) variability. They both show randomly distributed peaks in the count rate at the same time. The V light curve is  much less variable. We note that the recorded peaks are independent from the orbital period which is only $\sim15.1$~h \citep{chevalier89}.

We calculated the fractional root mean square variability, F$_{\rm var}$ \citep[][]{vaughan05}, for the light curves in Fig. \ref{fig:one}.  
All uncertainties are hereafter at the $1\sigma$ confidence level.
F$_{\rm var}$ is $73.0\pm1.5\%$ for the X-ray band,  $50.0\pm1.4\%$ for the UVW1 band, and $10.0\pm1.6\%$ for the V band, implying that the X-ray emission is the most variable. The most significant changes in the X-ray count rate are detected between 39.5 d and 43.3 d and between 97.2 d and 101.5 d,  where the count rate respectively decreases by a factor of $\sim13$ and increases by a factor of $\sim22$ in four days only (from $0.21\pm0.01$ c/s to $0.016\pm0.04$ c/s, and from $0.006\pm0.003$ c/s to $0.13\pm0.01$ c/s respectively).

\subsection{Structure function}
\label{struc}

\begin{figure*}
\begin{center}
\begin{tabular}{cc}
\includegraphics[angle=-90,width=3.0in]{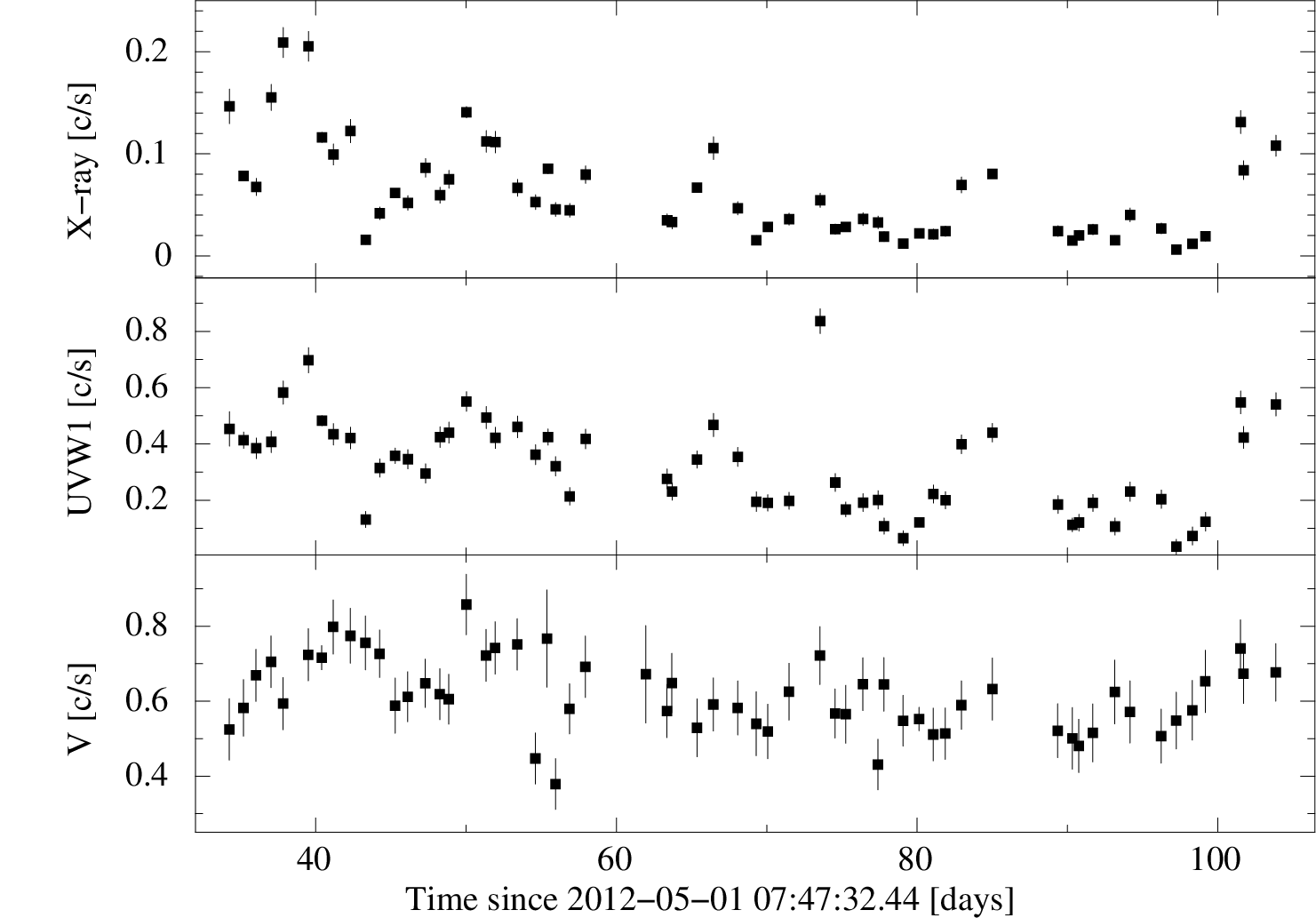} &
\includegraphics[angle=-90,width=2.8in]{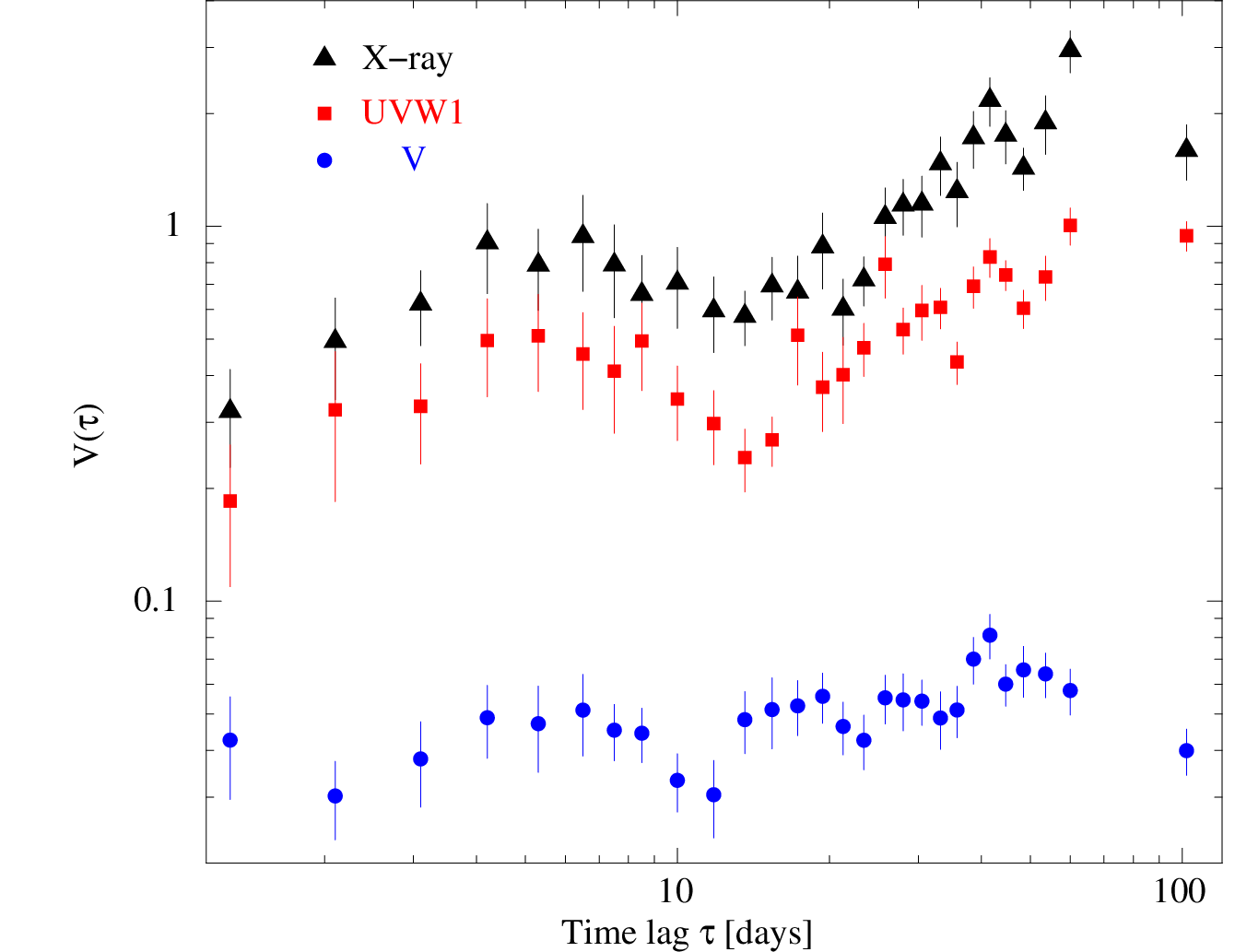} \\
\includegraphics[angle=-90,width=3.0in]{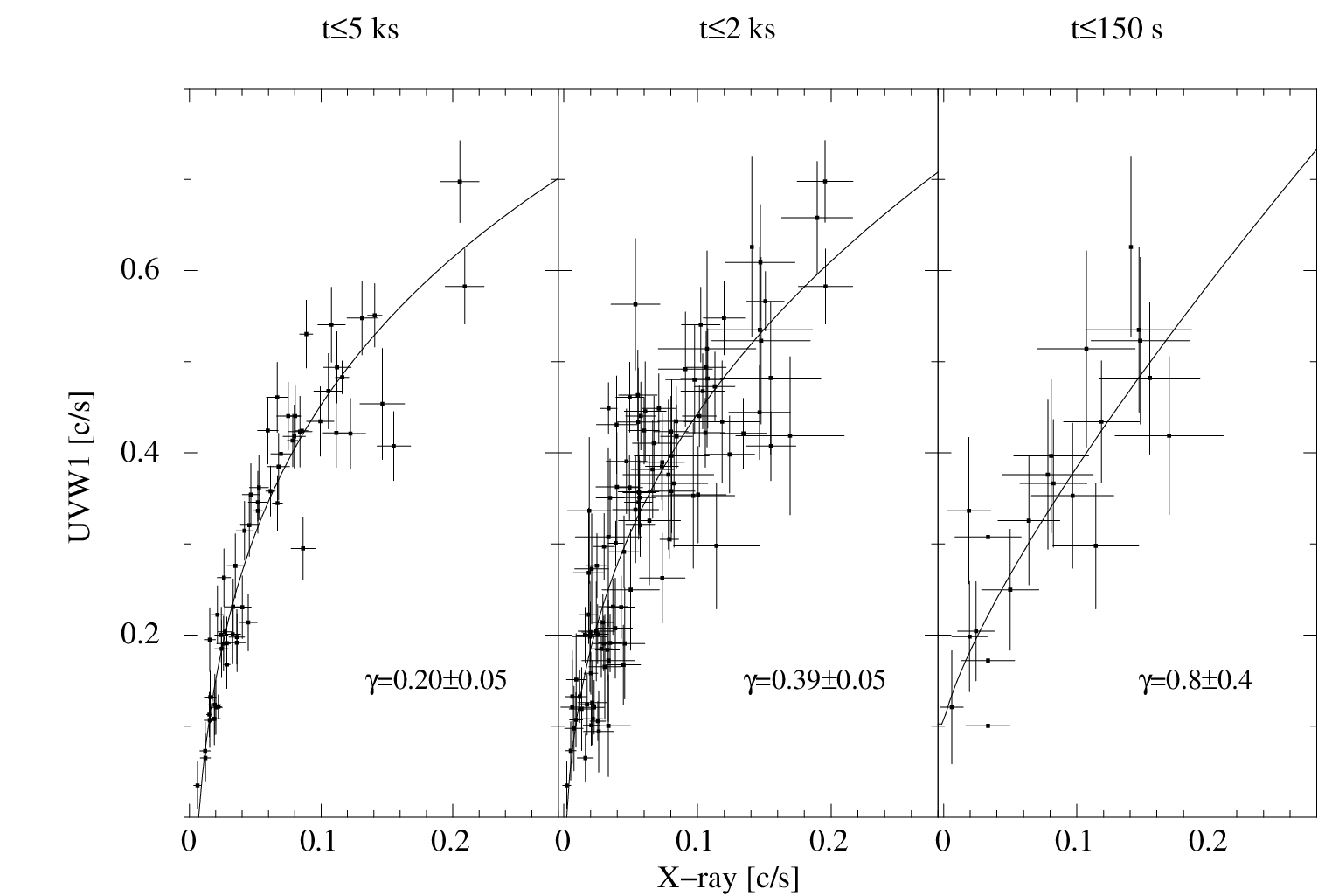} &
\includegraphics[angle=-90,width=3.0in]{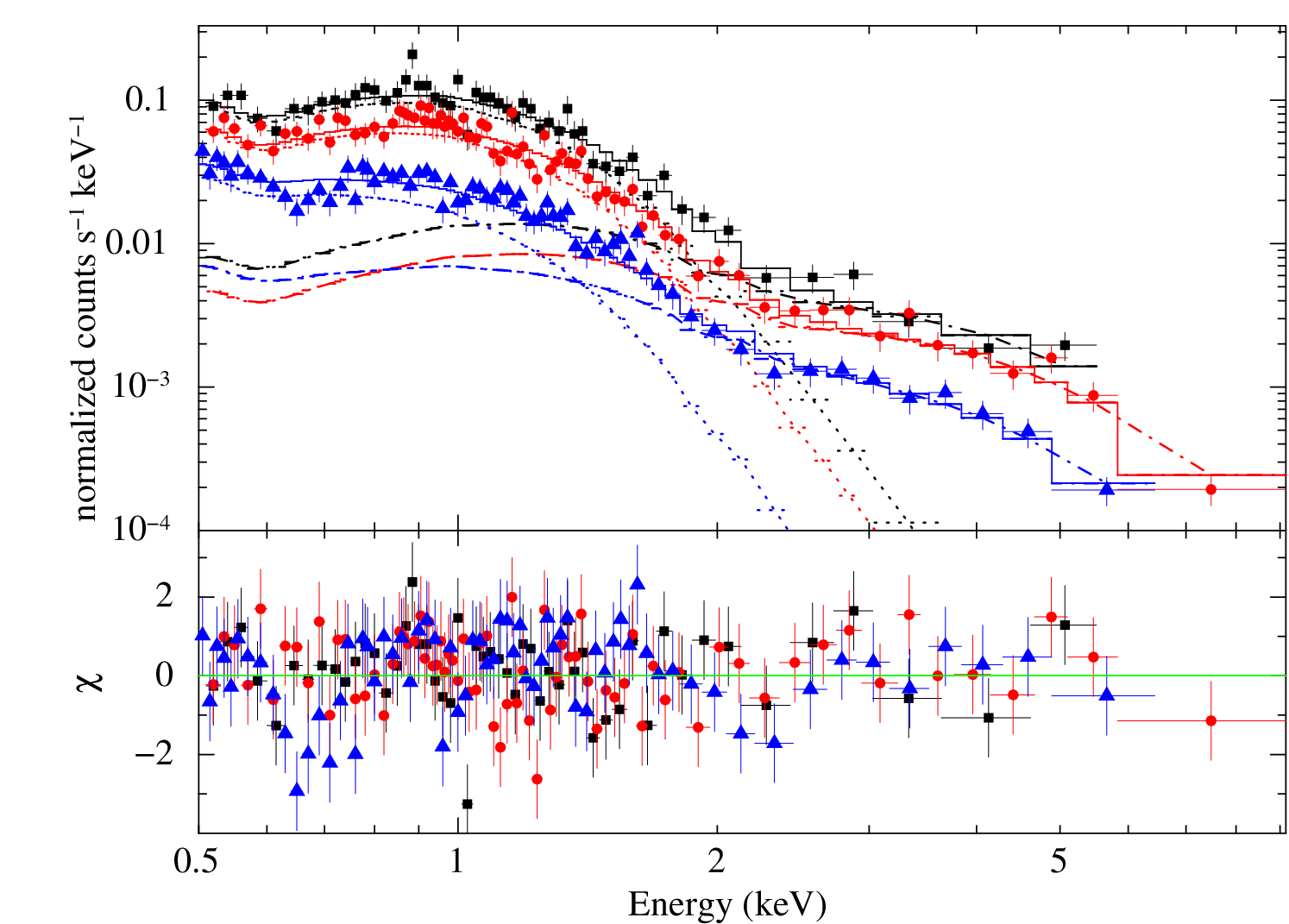} \\
\end{tabular}
\caption{{\it Upper Left}: Cen X-4 light curves in three energy ranges. 0.3--10 keV, X-ray (top),  
UVW1, 2600\AA\ (center) and V, 5468\AA\ (bottom). {\it Upper Right:} First order structure function $V(\tau)$ as a function of the time lag for the three light curves (X-ray, UVW1, and V) in the left panels. The light curves have been normalized to their average value before computing the structure function. This allows for easy comparison of the variability amplitude in each band. {\it Lower Left:} Study of the X-ray vs UV and optical correlation (only the UVW1 band is shown) as a function of the timescale. The timescale decreases from left to right. The solid line represents the best fitting model, a constant plus a power law. The value of the slope $\gamma$ is also reported. X-ray vs UV-O count rate on: the XRT timescale, t$\leq5$ ks (left), the UVOT snapshot timescale, t$\leq2$ ks (center) and the UVOT shorter snapshot timescale t$\leq150$ s (right). \textit{Lower Right:}  Cen X-4 summed spectra for three count rate ranges: low $<0.07$ c/s (blue triangles), medium 0.07--0.11 c/s (red circles), high $>0.11$ c/s (black squares). The solid line represents the total model, while the dotted line represents the \textit{nsatmos} component and the dash dotted line the power law component. Model residuals are showed in the lower panel.}
\label{fig:one}
\end{center}
\end{figure*}

We compute the structure function, $V(\tau)$ \citep[][]{do09}, for the three light curves (X-ray, UVW1, and V) in Fig. \ref{fig:one}. 
The slope of the structure function is related to the slope of the power spectrum.  For instance, a power spectrum with $P(f) \propto f^{-2}$ will show a first-order structure function where $V(\tau) \propto \tau^{1}$ \citep[][]{hughes92}. We find the power law index of the structure function of Cen X-4 to be $1.0\pm0.1$ and $0.8\pm0.1$ and $0.3\pm0.1$ in the case of the X-ray, UVW1, and V respectively (see Fig. \ref{fig:one}). The slope of the X-ray and UVW1 band is consistent with a power spectrum with index $-2$ as typically seen in accreting systems, while the slope of the V band is different and, consequently, it may be associated with another underlying process. The structure function also shows that the amplitude of the variability is greater in the X-ray band than in others, and the UVW1 amplitude is greater than in the V band. The timescale of the variability goes from 10 days up to at least 60 days (where the power law extends). However, an excess is present at about 4-5 days. This is likely linked to the rise and decay time of the peaks in the light curve.

\subsection{Correlation}

We found a highly significant ($\sim8\sigma$) and strong correlation between the X-ray and the UV (UVW2, UVM2, UVW1, U) emission on long timescale (t$\leq5$ ks), on short timescale (t$\leq2$ ks), and also on very short timescale (down to t$\leq$110 s).  We found a significant but less intense correlation also between the X-ray and the V band on long timescale (t$\leq5$ ks), and on short timescale (t$\leq2$ ks). The most likely shape of the correlation is a power law with index $\Gamma=0.2-0.6$ (see Fig. \ref{fig:one}). 

\subsection{X-ray spectral analysis}

We produced three averaged X-ray spectra as a function of the source count rate, low $<0.07$ c/s, medium 0.07--0.11 c/s, and high $>0.11$ c/s. We found the spectra well fitted by a model made by the sum of a thermal component in the form of hydrogen NS atmosphere plus a non-thermal power law tail, both multiplied by the interstellar absorption. Both components must change with the flux. The overall spectral shape remains the same with respect to flux variability. Moreover, we also compared, for each XRT observation, the 0.3--2 keV count rate with that in the 2--10 keV energy band. A linear model perfectly fits the data (not shown). This confirms that the overall spectral shape remains constant as the flux changes.
The thermal and the power law component are changing in tandem, being both responsible always for the same fraction, about 50\%, of the total X-ray flux in the 0.5--10 keV band. This suggests that they are physically linked. The temperature of the NS surface is changing being hotter when the flux is higher: kT$_{h}=80\pm2$ eV, kT$_{m}=73.4\pm0.9$ eV, kT$_{l}=59.0\pm1.5$ eV (while $\Gamma_{h}=1.4\pm0.5$, $\Gamma_{m}=1.4\pm0.2$, $\Gamma_{l}=2.0\pm0.2$). We conclude that accretion is very likely still occurring also at low Eddington luminosity rates ($\sim10^{-6}$) and that the matter is reaching and heating the NS surface, overtaking the action of the centrifugal force of the rotating magnetosphere. Since the two spectral component are closely linked, accretion is likely generating somehow also the power law component.

\section{Discussion}
\subsection{Evidence for accretion}

We showed that the X-ray and UV light-curve of Cen X-4 during quiescence are both highly variable, with the fractional root mean square variability F$_{\rm var} = 73.0\pm1.5\%$ and $50.0\pm1.4\%$ respectively, while the optical light curve is changing less (F$_{\rm var}=10.0\pm1.6\%$). The first order X-ray and UVW1 structure function ($V_{\tau}$) shows the shape expected for a power spectrum with index -2, as is typically seen in accreting systems. The timescale of the variability underlying the structure function goes from few days ($\sim5$) up to months. In the time between different peaks the count rate seems to always reach a minimum (ground) level extremely close to the detection limit for both the X-ray and the UVW1 band. We observed the strongest X-ray variability on short timescale ever detected for Cen X-4 during quiescence a factor of 22 and a factor of 13 in only four days.

Moreover, we showed that both the spectral components, the NS atmosphere and the power law, must change with the flux.  We also showed that they are changing in tandem, implying that they are closely linked. A change in the thermal component that we detected as a surface temperature change strongly suggests that the accreting matter must reach the surface of the NS, overtaking the centrifugal barrier of the rotating magnetosphere. Since the two spectral component are so closely linked, accretion must be, very likely, also responsible for the change in the power law. A very low level of accretion seems to occur quite continuously, generating random episodes of high count rates and consequently modifying the temperature of the surface, which is hotter when the count rate is higher. We note that it is extremely difficult to account for such intense, non monotonic, multi-flare-like variability, with a clear correlation between the thermal and power law spectral component, without invoking accretion that is somehow occurring at very low Eddington luminosity rates in Cen X-4.

\subsection{Origin of the UV emission}

In quiescent LMXBs the donor stars are considered intrinsically too cool to have any significant UV emission \citep[][]{hynes12}. The UV, then, could probe the emission from a particular region of the accretion flow. However, the exact location of origin of the UV emission in the accretion flow is still unclear and it is consequently debated. The UV emission could arise due to:  (I) the thermal emission from the gas stream-impact point,  (II) the thermal emission from a standard, optically thick and geometrically thin accretion disc truncated far from the compact object, (III) the emission from an advection dominated accretion flow, ADAF \citep[][]{narayan1,narayan2}, or alternatively the UV emission could be due to (IV) X-ray irradiation of, and reprocessing (perhaps in the inner accretion disc or on the surface of the companion star).

\subsubsection{Thermal emission from the gas stream-impact point}

Our analysis clearly showed that the UV and the X-ray emission are strongly correlated.
The presence of a correlation, especially that at very short timescales (110 s) clearly rules out the thermal emission from the stream impact point as source of the UV photons. Indeed, if we assume that the UV light is tracing the matter in the disc, UV variations would refer to instability propagation throughout the disc. However, the timescale to traverse the whole disc from the stream impact point to the inner edge is significantly longer than the correlation timescale, being of the order of several weeks \citep{fkr92}. Such a long timescale cannot explain the correlation we are observing here.

\subsubsection{Thermal emission from a truncated standard accretion disc}
\label{subsub:thermal}

In the standard disc solution the radial velocity of the matter in the disc is of the order of 0.3 km s$^{-1}$ \citep{fkr92}, and the measured timescale of the correlation is t$\leq110$ s (U band). Consequently, if the observed correlation is generated by matter propagating in a standard disc, finally accreting on the NS surface, this matter must propagate from a maximum distance of about 33 km from the NS center. However, this would imply that the disc is extending down to the NS surface which is not observed (the X-ray thermal emission is consistent with the emission from the NS surface only).  Moreover, if we impose that the viscous timescale is equal to the shortest correlation timescale, we could get an estimate of the size of the disc in an extremely peculiar scenario, where the dynamics in the disc would be fast enough to justify the observed correlation. By using t$_{c}\leq110$ s we derive $R\lesssim9$ km, a totally non physical value, as it is smaller than the size of the NS itself. 
We conclude that a standard disc cannot account for the observed correlation, implying that UV variability does not arise from instabilities propagating through a standard disc.

\subsubsection{Advection dominated accretion flow, ADAF}
\label{subs:corrad}

We can estimate the maximum distance to the NS from which the matter is accreting in an ADAF (R$_{acc}$), assuming that the accretion timescale is equal to the shorter correlation timescale we found ($t_{cor}\leq110$ s, for the V band). This is a natural assumption since only if $t_{acc}<t_{cor}$ can accretion via an ADAF be the cause of the observed correlation. Since $R=(t_{acc}\,0.1)^{2/3}(GM)^{1/3}$, for $M=1.4M_{\odot}$ we get R$_{acc}\lesssim6200$ Schwarzschild radii, in agreement with theoretical values \citep[][]{narayan2}. Consequently, the ADAF scenario can account for the timescale of the correlation. The variability in the amount of matter that the star is accreting from a region inside of a radius of $R\sim6200R_{s}$ could trigger X-ray variability. However, with present spectral data, we were unable to verify if  the ADAF model can accurately reproduce the whole source SED from optical to X-ray (not shown). Summarizing, while the ADAF alone is very promising to explain the timescale of the observed accretion, however, it seems to be unable to explain the intensity and the peak of the UV emission without invoking some still not well understood extra effect.

\subsubsection{X-ray irradiation and reprocessing}

According to the reprocessing scenario the UV flux has to be only a small fraction of the X-ray one. The maximum fraction of the reprocessed emission $S$ has to be equal to the fraction ($S^{'}$) of the sphere surface of radius $R$ occupied by that part of the disc reprocessing the X-ray radiation. $S=S^{'}$ only works if the reprocessing is $100\%$ efficient, otherwise $S<S^{'}$. We emphasize that for a standard disc, $S^{'}$ is always extremely small, as the disc is geometrically thin. For Cen X-4 $S^{'}\leq0.8\%$. However, the value of $S^{'}$ does not match the amount of UV reprocessing estimated from the observation, where $S=F_{UV}/F_{X}$ is $2.1\%<S<23\%$. We therefore conclude that $2.1-23\%$ is not a reasonable value for reprocessed emission from a standard thin disc.
On the contrary, we note that the fraction of the sphere of radius $R$ occupied by the surface of the companion star, is consistent with the observed fraction of reprocessed light. In this case $2.3\%<S^{'}<3.4\%$. We conclude and stress that the companion star is the most likely source of reprocessing, while the accretion disc alone cannot.

Moreover, the SED (not shown) pointed out the presence of a small hot spot ($\sim16000$ K, $\sim2.4\times10^{4}$ km, fractional area of about $2\%$) on the companion star, generated by X-ray irradiation, likely in a region where no shield is provided by the accretion disc. We also found that the accretion disc must be intrinsically UV faint, as its contribution to the total UV light is less than $3\%$.

Finally, we emphasize that the ADAF and reprocessing scenario are not mutually exclusive, they in fact could easily coexist. Indeed, if most if not all the UV emission is produced by reprocessing from the companion star, as we have shown, it is not surprising that the ADAF model alone can not account for the intense emission in the UV band as suggested by \cite[][]{menou01}.  We note that, at least for what concerns Cen X-4, a detailed model including ADAF plus reprocessing from the companion should be tested against the data.

\end{document}